# Molecular Mechanism Enabling Linearity and Symmetry in Neuromorphic Elements

Bidyabhusan Kundu and Sreetosh Goswami*
Centre for Nano Science and Engineering, Indian Institute of Science, Bangalore
Email id: sreetosh@iisc.ac.in

**Abstract:**

For over a decade, linear and symmetric weight updates have remained the elusive holy grail in neuromorphic computing. Here, we unveil a kinetically controlled molecular mechanism driving a near-ideal neuromorphic element, capable of precisely modulating conductance linearly across 16,500 analog levels spanning four orders of magnitude. Our findings, supported by experimental data and mathematical modelling, demonstrate how nonlinear processes such as nucleation can be orchestrated within small perturbation regimes to achieve linearity. This establishes a groundwork for routinely realizing these long-sought neuromorphic features across a broad range of material systems.

**Main text:**

Dot product engine (DPE) is a neuromorphic accelerator the foundation of which trace back to Carver Mead's work in the 1980s. It can dramatically simplify vector-matrix multiplication — a fundamental mathematical operation that underpins many computational tasks, from signal and image processing to machine learning, scientific computing, and artificial intelligence. In contrast to traditional digital computers, which require $n^2$ computational steps to perform matrix-vector multiplication ($O(n^2)$ complexity), the DPE can, in principle, achieve this in a single step, reducing the computational complexity to $O(1)$ (Fig. 1a)[1-3]. However, the practical application of DPEs, has been hindered by persistent challenges that went unsolved for over a decade.

First, the resolution of typical DPEs, ranging from 2 to 6 bits, limits their utility to a narrow set of computing problems, further exacerbated by the formation of sneak paths within the crossbar architecture. Secondly, the process of writing data into the crossbar is challenging due to the nonlinear and asymmetric nature of weight updates of typical neuromorphic elements. The nonlinearity causes the required pulse trains to vary considerably depending on the initial and target weight values, introducing substantial computational overhead. Moreover, the inherent variability in weight update characteristics necessitates frequent error correction techniques, such as write-and-verify loops, which can take several milliseconds for each crosspoint[4]. This not only slows down the data-writing process but also renders it highly energy-inefficient, and depletes the endurance cycles of individual crosspoints, considerably diminishing the advantages of a DPE.

Solution to these challenges has been well-articulated and extensively documented in the literature[1-3,5]. An analog circuit element capable of storing and processing data across thousands of conductance levels, with linear and symmetric updates, could give a one-shot solution to almost all of these problems. The primary challenge has been in identifying a material and the suitable transition mechanism that could enable such an ideal analog response.

Conventional memristive materials mostly employed in DPEs are oxides. There analog levels arise from the gradual formation and rupture of conductive filaments comprised of oxygen vacancies or metal ions diffused from the electrodes[6-8]. The ion migration processes, governed by the local electric field (*i.e.*, drift), concentration gradients (*i.e.*, diffusion), and temperature variations, are inherently nonlinear, anisotropic, and they differ across various conductance levels[9-11]. Consequently, the weight update process substantially deviates from linearity. The clustering of ions during the write operations are stochastic that also affect the reliability of the weight update (see Fig. 1b). Notably, similar challenges also affect linearity in nitride films, perovskites[12-14] and two-dimensional (2D) material-based memristors[15-17] (and memtransistors), which also rely on ion migration.

Phase-change memory (PCM) is another mature technology utilized in DPEs that hinges on a reversible phase transition between an amorphous (high-resistance) state and a crystalline (low-resistance) state, induced by electric pulses that trigger local heating[18-20]. The temperature distribution within the PCM is non-uniform, causing disorder in the atomic structure of the amorphous region during the melt-quench processes making the process inconsistent and non-linear. The crystallization process is also influenced by precursor sites serving as nucleation centres within the amorphous regions that introduce stochasticity, noise and drift (Fig. 1b).

Several emerging materials and technologies have also been explored in pursuit of an optimal neuromorphic element. A notable example is ferroelectric materials (perovskites, 2D, molecular), incorporated into various device architectures such as ferroelectric tunnel junctions (FTJs)[21-24], and ferroelectric field-effect transistors (FeFET)[25,26]. The switching mechanism in ferroelectric materials is determined by gradual change in polarization, driven by an applied voltage pulses. These are nucleation-limited processes comprising slow domain nucleation followed by rapid domain expansion and eventual domain merging, all of which are inherently nonlinear. Electrochemical synapses, another promising class of biomimetic platforms that have recently garnered attention, operate through charge injection. Since these are inherently non-equilibrium structures, they are fundamentally nonlinear[27-29]. Although efforts have been made to confine conductance windows to achieve quasi-linear behaviour, these restrictions often reduce the number of accessible states, thereby limiting the precision of dot product engines and diminishing their overall utility (see Fig. 1b).

In this work, we demonstrate that voltage-driven molecular kinetics within a small perturbation regime enable perfectly linear and symmetric weight updates across four orders of conductance, providing reliable access to 16,520 distinct analog levels (Fig. 1c). This near-ideal neuromorphic behaviour is driven by a zeroth-order molecular kinetic process. Fig. 2a shows the Ru complex of a bis-azo-aromatic ligand used as the molecular memristive material[3]. Fig. 2b shows its ground state orbital energy levels[30]. Different electronic states of this molecule, obtained under the application of voltage are shown in Fig. 2c. The electronic states, labelled as 00, 11, 22 and 31 represent the occupancy of the antibonding orbitals in a pair of adjacent molecules. '0' denotes an unreduced molecule, whereas 1, 2 and 3 indicate molecules with one, two and three additional antibonding electrons, respectively. Notably, while the redox states (00), (11) and (22) are symmetric, (31) is a symmetry broken charge disproportionate (CD) state with two adjacent molecules, one triply and the other singly reduced. These states were characterized by *in-operando* Raman and UV-Vis spectroscopy, as discussed in ref.[3,30,31]. Notably, the CD state is stabilised by the coulombic gradient resulted from counterion displacement around the molecules. The ion-displacement involved in a symmetry breaking process is about 40-times (computed from quantum chemical calculations) higher than the other redox transitions (ref.[30]) (Fig. 2d). In Fig. 2e we compare the intermolecular coupling strengths of the electronic states depicted in Fig. 2c, which govern the conductance (*i.e.*, the electron transfer function) through the molecular film in each state. The coupling strength is highest in the 00 and 22 states, attributed to a uniform electron cloud within the coordination sphere. The 11 state shows moderate coupling, while the 31 state demonstrates the weakest coupling due to electronic inhomogeneity caused by symmetry breaking[30].

Fig. 3 compares the steady-state (or quasi-static) *I(V)* response with pulsed measurements obtained from the molecular film depicted in Fig. 2. The measurements were performed in a crossbar structure, shown in Fig. 1a. The two sets of characteristics differ strikingly. In both cases, the ground state of the film is 31[3]. In the steady-state measurements, a sharp transition to the 11 state occurs at around 2 V ($V_1$), followed by another abrupt transition to the 00 state at 2.76 V ($V_2$), as shown in Fig. 3a. In contrast, when subjected to an 80 ns, 900 mV pulse train (Fig. 3b), the film undergoes a gradual transition to the 22 state — a molecular configuration never observed in steady-state measurements.

The molecular states described above were attributed based on in-situ Raman spectroscopy. The pseudo-colour plot in Fig. 4a captures two prominent sharp transitions: from 31 to 11 (at $V_1$) and subsequently from 11 to 00 (at $V_2$), as also illustrated in the reaction coordinate diagram in Fig. 4c. Fig. 4b highlights the gradual transition from 31 to 22 under pulsed conditions, which is also depicted in the reaction coordinate diagram in Fig. 4c. We inferred that during steady-state measurements, the millisecond duration of applied voltages enables complete thermodynamic transitions (Fig. 3a), leading to sharp switching. In contrast, the 80 ns pulse duration in pulsed measurements is insufficient to transform the entire film volume (Fig. 3b). Hence, the transition becomes kinetically limited, with each pulse knocking only a fraction of the total volume (Fig. 4c).

To further understand the difference between the steady-state (quasi-DC) and pulse measurements, we conducted time resolved conductance measurements combined with *in-operando* spectroscopy (Fig. 5). We applied a 1.2 V pulse of >2 μs duration that changed the film conductance by about six orders of magnitude (from point i to point iii, as shown in Fig. 5a). Note that these transitions are non-volatile. To confirm the molecular state at point iii, we removed the pulse and performed Raman spectroscopy. The spectrum precisely matched the signature of the 22 state, characterized by a single azo-stretching mode at 1259 cm$^{-1}$ as shown in Fig. 5b (also see ref.[3]). However, on applying the same 1.2V pulse for milliseconds, the 22-state gradually transitioned to a mixed, non-conducting state comprising 22 and 11 (points iv, v and vi) which was also spectroscopically measured (Fig. 5b). Subsequently, on applying a 2V pulse the film transforms to 11 state (point vii in Fig. 5a, b). This measurement sheds light on the effect of time scale of measurement on the electronic transitions. In steady state measurements, each voltage value was stable for >50 ms (see Fig. 3b), which is much longer than the full measurement timescale shown in Fig. 5a. Thus, the pure 22 state (which is conducting) is non-observable in steady state measurement which transforms to non-conducting (22+11) mixed state. From the temporal dynamics, we conclude that at 1.2V, the 22 and 11 states are energetically degenerate (Fig. 5c), as evidenced by the equipartition mixture of these states at point vi in (Fig. 5a, b). However, transition to a pure 11 state requires overcoming a significant energy barrier[32,33]. Quantum chemical calculations indicate that the energy barrier between the 31 and 22 states is only 8% of the barrier between the 31 and 11 states. To further test this hypothesis, we lowered the temperature to 190K at point iii in Fig. 5a. At this reduced temperature (points viii, ix, and x), the 22 state became frozen, preventing the formation of the mixed state (22+11) observed at higher temperatures. Upon reheating at point x, with the applied voltage maintained at 1.2V, the mixed state (22+11) reappeared, confirming the kinetically limited transition depicted in Fig. 5c. Applying a 2V pulse lowered the kinetic barrier (Fig. 5c) sufficiently to drive the transition to the 11 state, as observed at both 294K and 190K, explaining the sharp transitions seen in DC sweep measurements. Note that with 900mV pulses of 80 ns width, the barrier between the 22 and 11 states could not be overcome, resulting in access to either the 31 state or a mixture of 22+31 or the 22 state, which explains the observed analog levels.

These temporal dynamics observed in Fig. 5 stem from the movement of counterions around the molecules (Fig. 2d). As illustrated in Fig. 6a-c, The transitions to and from the 31 state involve two processes, *viz.* ion displacement and electronic rearrangement. Note that the electronic transition between the 31 and 22 states conserves total charge, ensuring no net charge transfer to or from the electrode. Consequently, the system retains a charge equilibrium state, unlike electrochemical transistors where charge injection is inherently non-equilibrium and nonlinear[6,7,27,28].

Moreover, the electronic transition and ion displacement operate on entirely different timescales. Electronic processes are fast (typically sub ns- ref.[34,35]) while ionic movements are slower due to their vastly larger reduced masses[36]. As a result, these molecular transitions can effectively be viewed as two-step processes (see refs.[30,31]): a rapid intermolecular electron transfer coupled with a slower, kinetically limited counterion relaxation along the resulting Coulombic gradient (see Fig. 6 a-c). The ion displacement imparts non-volatility to the transitions by stabilizing the relaxed state through multipathway (or many-body) molecular interactions (ref.[31,37]).

**Table 1 –** Different configurations of molecular redox states and the corresponding counterion positions.

| LABEL | MOLECULAR STATE |
|---|---|
| $31_{31}$ | Electronic state 31, and the counterions have relaxed to their optimal pockets to stabilize the 31 state |
| $22_{31}$ | Electronic state 22, but the counterions still in their optimal pockets attained in 31. This species is an intermediate between $31_{31}$ to $22_{22}$ transition |
| $22_{22}$ | Electronic state 22, and the counterions have relaxed to the optimal pockets to stabilize the 22 state |
| $31_{22}$ | Electronic state 31, but the counterions still in their optimal pockets attained in 22. This species is an intermediate between $22_{22}$ to $31_{31}$ transition |

In Fig. 6a and b, we label distinct molecular configurations as $31_{31}$, $31_{22}$, $22_{31}$, and $22_{22}$, where the subscripts denote the respective counterion positions as depicted in Fig. 6c and described in Table 1. During potentiation, the $31_{31}$ state initially transitions to $22_{31}$ (Fig. 6a, c) via rapid electron transfer. The $22_{31}$ state is volatile, and depending on the applied pulse width, a fraction undergoes counterion relaxation forming the non-volatile $22_{22}$ state, while the remainder reverts to $31_{31}$. This description is consistent with Raman measurements shown in Fig. 4. The depression process follows a similar mechanism, as shown in Fig. 6b. The electronic transitions occur too rapidly to measure experimentally, whereas the slower ionic transitions determine the overall rate. By analysing the temperature-dependent rates of transitions between consecutive conductance levels (Fig. 6d), we plotted the logarithm of transition rates (log R) against the inverse temperature ($1/(k_B T)$). The overlapping data points for various analog transitions (e.g., 1-2, 5,001-5,002, and 16,001-16,002) indicate identical activation energy ($E_a$) and attempt frequency ($\nu$), confirming that the rate of the analog transitions remain constant, *i.e.*, they follow zeroth-order kinetics, which corroborates the observed linearity. The activation energy ($E_a$) of approximately 130 meV agrees with quantum chemical calculations (Fig. 6e), which reveal a volumetric expansion indicative of ion relaxation through displacement.

To establish a design framework for obtaining materials with linear and symmetric weight update characteristics, we constructed a mathematical framework to quantitatively analyse electron transport in our devices. This model enabled us to define a parameter space governing the dynamics of the weight update process. As discussed in previous literature, a simple resistor network model cannot explain electron transport in molecular systems[38-40]. We formulated an N-site hopping mechanism, using a 1-D transfer function along the z-axis and incorporating areal inhomogeneity along the x-y plane. Fig. 7a illustrates the device structure and Fig. 7b captures the abstraction used in our model. We assumed a molecular chain with $N$ = 30 hopping sites, corresponding to thirty molecular layers along the z-direction in a 60 nm film. Each layer represents a pair of molecules, with a thickness around 2 nm, consistent with quantum chemical calculations (see Fig. 7a).

We modelled the current, the rate of electron flow, using rate equations comprising interfacial and intermolecular hopping rate constants that allowed us to calculate the occupation probabilities at various hopping sites[39-42]. In this formulation, the forward and backward electron transfer rates between the top and bottom electrodes and the molecular layer are denoted as $K_{T,f/b}$ and $K_{B,f/b}$, where T and B refer to the top and bottom electrodes, and f and b denote forward and backward directions, respectively. These rates are proportional to the electrode-molecule coupling $\Gamma_T$ and $\Gamma_B$ for the top and bottom electrodes, respectively. The intermolecular electron transfer rate constant, $K_{m,f/b}$, where m denotes the molecule and $f$ and $b$ refer to the forward and backward directions, is governed by the molecule-molecule coupling strengths $\gamma_m$. The various rate constants, illustrated in Fig. 7b, can be expressed as:

$$K_{m,f}(n) = \gamma_m(n) \times \exp\left[\frac{(a_m \times V_r)}{\eta k_B T}\right] \quad (1)$$

$$K_{m,b}(n) = \gamma_m(n) \times \exp\left[\frac{-(a_m \times V_r)}{\eta k_B T}\right] \quad (2)$$

$$K_{T,f}(n) = \frac{\Gamma_T(n)}{\sqrt{4\pi u_\lambda k_B T}} \times \int \exp\left[-\frac{(-u_\lambda - \Delta E - (a_T \times V_r) + E)^2}{4 u_\lambda k_B T}\right] \cdot \psi(E) \cdot dE \quad (3)$$

$$K_{T,b}(n) = \frac{\Gamma_T(n)}{\sqrt{4\pi u_\lambda k_B T}} \times \int \exp\left[-\frac{(-u_\lambda + \Delta E - (a_T \times V_r) - E)^2}{4 u_\lambda k_B T}\right] \cdot (1 - \psi(E)) \cdot dE \quad (4)$$

$$K_{B,f}(n) = \frac{\Gamma_B(n)}{\sqrt{4\pi u_\lambda k_B T}} \times \int \exp\left[-\frac{(-u_\lambda + \Delta E + (a_B \times V_r) - E)^2}{4 u_\lambda k_B T}\right] \cdot (1 - \psi(E)) \cdot dE \quad (5)$$

$$K_{B,b}(n) = \frac{\Gamma_B(n)}{\sqrt{4\pi u_\lambda k_B T}} \times \int \exp\left[-\frac{(-u_\lambda - \Delta E - (a_B \times V_r) + E)^2}{4 u_\lambda k_B T}\right] \cdot \psi(E) \cdot dE \quad (6)$$

Where $\Delta E$ is the hopping energy gain and $u_\lambda$ is the interfacial re-organization energy, $n$ is the index in the conductance level ranging from 1 to 16,500 (accessed by $n$ number of 900mV, 80ns pulses). In any electronic device, a fraction of the pulse amplitude $a_T \times V_r$ and $a_B \times V_r$ would drop across the top and bottom electrode respectively and the rest $a_m \times V_r$ appears across the molecule. Here $V_r$ is the reading pulse amplitude, and $a_T + a_B + a_m = 1$. $\eta$ is a fitting parameter derived from variations in the measured current corresponding to changes in the $V_r$. $\psi(E) = 1/[1 + exp(E/k_B T)]$ is the acceptor density of states at energy $E$, where $k_B$ is the Boltzmann constant and $T$ is the operating temperature.

The occupational probabilities at the different hopping sites $P_i$ in the stationary limit are given by,

$$\dot{P}_1 = 0 = -(K_{L,b} + K_{m,f})P_1 + K_{m,b}P_2 + K_{L,f}P_{T,B} \quad (7)$$

$$\dot{P}_2 = 0 = K_{m,f}P_1 - (K_{m,b} + K_{m,f})P_2 + K_{m,b}P_3$$

$$\vdots$$

$$\dot{P}_{N_z} = 0 = K_{m,f}P_{N-1} - (K_{m,b} + K_{R,f})P_N + K_{R,b}P_{T,B}$$

To ensure no charge accumulation, the normalization condition is expressed as,

$$P_1 + P_2 + P_3 \ldots + P_N + P_{T,B} = 1 \quad (8)$$

The occupation probabilities are determined by solving the $N + 1$ linear equations (eqn. 7 and 8). From these equations, the current $I(n)$ in the conductance level $(n)$ ranging from 1 to 16500, can be calculated as:

$$I(n) = e\left(K_{B,f}(n)P_{T,R}(n) - K_{B,b}(n)P_1(n)\right) = e\left(K_{m,f}(n)P_{N-1}(n) - K_{m,b}(n)P_N(n)\right) =$$
$$e\left(K_{T,f}(n)P_N(n) - K_{T,b}(n)P_{T,B}(n)\right) \quad (9)$$

As the film molecules gradually transition between the 31 and 22 states (or vice versa), multi-pathway interactions among the molecules and ions[31,37] influence the energy levels and couplings, thereby impacting all the rate constants ($K_{T/B/m,f/b}$) in equations 1-9. The coupling strengths are governed by π-orbital delocalization at a specific electronic state. Basic quantum calculations[43-45] indicate that molecular coupling is proportional to the broadening observed in the energy levels. To assess this, we measured the full width at half maximum (FWHM) of the Raman mode at 1259 cm$^{-1}$, corresponding to the stretching of a singly reduced azo bond that hosts the LUMO $+x$ (where $x$ in an integer and $x \in [1,4]$) orbitals for electron hopping. As shown in Fig. 7c, the FWHM of this azo mode increases linearly with the fraction of 22-states in the film, which is also proportional to the pulse count. Based on this proportionality, the coupling parameters in equations (1-6) were formulated as follows:

$$\gamma_m(n) = \left[\left(\sum_1^n df_{31}(n)\right) \times \gamma_{m,31}\right] + \left[\left(\sum_1^n df_{22}(n)\right) \times \gamma_{m,22}\right] \tag{10}$$

$$\Gamma_T(n) = \left[\left(\sum_1^n df_{31}(n)\right) \times \Gamma_{T,31}\right] + \left[\left(\sum_1^n df_{22}(n)\right) \times \Gamma_{T,22}\right] \tag{11}$$

$$\Gamma_B(n) = \left[\left(\sum_1^n df_{31}(n)\right) \times \Gamma_{B,31}\right] + \left[\left(\sum_1^n df_{22}(n)\right) \times \Gamma_{B,22}\right] \tag{12}$$

Here, $df_{22}(n)$ represents the fractional change in the 22-state between two consecutive conductance levels, $n$ and $n-1$, and is defined as $df_{22}(n) = f_{22}(n) - f_{22}(n-1)$. Using Marcus theory[46-49] and nucleation[50-54], we developed a mathematical framework for quantifying $f_{22}(n)$ and $df_{22}(n)$ (Fig. 8) which were then compared with the trends extracted from in-operando Raman spectroscopic measurements, shown in Fig. 4. $\gamma_{m,22}$ and $\gamma_{m,31}$ are inter-molecular coupling constant in 22 and 31-state respectively. Similarly, $\Gamma_{T/B,22}$ and $\Gamma_{T/B,31}$ are top/bottom electrode-molecule coupling constant in 22 and 31-state respectively. Note that, for simplicity, coupling constants are assumed to be same for the forward and backward directions.

Fig. 8a shows the potential profiles, $\Phi_{mol}(z,n)$, along the z-axis for different conductance levels, $n$, calculated using Thomas–Fermi screening framework (equations 13–16)[55-59]. Each layer in the device represents a pair of molecules in either the 31 or 22 state positioned between the top and bottom electrodes (Fig. 7). The charge-disproportionate 31 state exhibits a higher intermolecular dipole moment[30], which results in a greater screening ability compared to the 22 state. Consequently, in the 31 state, the potential profile within the device is relatively flat, with the potential drop occurring primarily near the molecule-metal contact. As the fraction of 22-state increases, the potential profile across the device becomes more linear. $\Phi_{mol}(z,n)$ was expressed as:

$$\Phi_{mol}(z,n) = \Phi_0(z) - \frac{V_m}{\pi} \sum_{m=1}^{\infty} \frac{F_m}{m(1+F_m)} \left[2 \times sin\left(m\pi\left(1 - \frac{z}{L}\right)\right)\right] \tag{13}$$

$$\Phi_0(z) = -\frac{z}{L} \times V_m \tag{14}$$

$$F_m = \frac{1}{2}\left(\frac{\sigma}{\zeta(n)}\right)^2 e^{\frac{1}{2}(2\pi m\sigma)^2} \int_{\frac{1}{2}(2\pi m\sigma)^2}^{\infty} du \frac{e^{-u}}{u} \tag{15}$$

$$\zeta(n) = [(b_1 n) + b_2] + [c_1 \times \exp(c_2 \times n)] \tag{16}$$

Here $\Phi_0(z,n)$ represents a linear potential profile which is obtained by solving the Laplace equation in the absence of any charge build-up inside the film. The parameter $\sigma$ is proportional to the van der Wall's radius of the molecules, normalized to the film thickness. $L$ is the thickness along z. Here $V_b$ is the writing pulse amplitude which is +900 mV and $V_m = a_m \times V_b$ ($a_m = 1 - (a_T + a_B)$) appear across the molecules. $\zeta$ represents the screening capability of the film, which is a function of conductance state ($n$). $b$ and $c$ are numerical constants.

Using $\Phi_{mol}(z,n)$ obtained from eq. 13, we calculated the electron transfer rate $k_{ET}(z,n)$ between the molecules based on eqns. 17-20, analogous to the Marcus-Hush-Chidsey integral[60,61]. Leveraging the multiple closely spaced LUMO levels in our molecular system, we constructed the following equations[46-49].

$$k_{ET}(z,n) = \frac{\gamma_{DA}}{\sqrt{4\pi E_\lambda k_B T}} \times \int \exp\left[-\frac{(E_\lambda + \Delta G^0 - e\Phi_{mol}(z,n) + E)^2}{4E_\lambda k_B T}\right] \cdot \psi(E) \cdot dE \quad (17)$$

$$\gamma_{DA} = \left(\frac{2\pi}{\hbar} \times H_{DA}^2\right) \quad (18)$$

$$P(z,n) = k_{ET}(z,n) \times \gamma_{DA}^{-1} \quad (19)$$

$$\theta(z,n) = \int_{-\infty}^{P(z,n)-\varepsilon} \delta(x)\,dx = \begin{cases} \theta(z,n) = 1, P(z,n) > \varepsilon \\ \theta(z,n) = 0, P(z,n) < \varepsilon \end{cases} \quad (20)$$

The transfer integral $H_{DA}$ represents the coupling between the donor and acceptor. For instance, during the 31→22 transformation, the 3 moiety acts as the donor, while 1 serves as the acceptor. $E_\lambda$ represents the reorganization energy required to move the counter-ions from their optimal 31 to the 22 positions. $\Delta G^0$ is the Gibbs free energy for the transition. $\Phi_{mol}(z,n)$ is the electrostatic potential at a given location $z$, and $\psi(E) = 1/[1 + exp\,(E/k_B T)]$ is the acceptor density of states at energy $E$, where $k_B$ is the Boltzmann constant and $T$ is the operating temperature. $dt = \gamma_{DA}^{-1}$ (equation 19), indicative of the timescale for the molecular transformation, normalizes $k_{ET}(z,n)$ to $P(z,n)$, *i.e.*, probability of electron transfer (see Fig. 8b, top panel). $\theta(z,n)$, calculated from $P(z,n)$ using eqn. 20[48,49], represents a parameter that takes discrete values of either 0 or 1, indicating whether a molecular transition from 31 to 22 has occurred at a specific $z$ and $n$ (Fig. 8b, bottom panel). Here, 0 corresponds to the 31-state (which is the ground state) and 1 to the 22-state. Where if $P(z,n) \geq \varepsilon$, the electron transfer process occurs and $\theta(z,n)$ switches from 0 to 1; otherwise, it does not. Notably, all the molecular parameters used here (*e.g.* $\Delta G^0$, $E_\lambda$) are consistent with quantum chemical calculations reported in ref.[3].

When $\theta(z,n) = 1$, the switching is energetically feasible only limited by the areal inhomogeneity along the $x - y$ plane (at various z positions) which we modelled using nucleation limited switching model as shown in eqn. 21 and 22 (see Fig. 8c):

$$Nr_{xy}(z,n) = \frac{1}{\pi} \times \arctan\left(\frac{n - \chi(z)}{\kappa}\right) + \frac{1}{2} \quad (21)$$

$$\chi(z) = a_1 \times z + a_2 \quad (22)$$

$Nr_{xy}(z,n)$ denotes the population of 22 along $x$-$y$ planes at different $z$-values (Fig. 8c). The transition of the entire film requires multiple nucleation events occurring at different $z$-positions depending on the $P(z,n)$ and $\theta(z,n)$ values (eqn. 19, 20). The rate of these nucleation processes is constant, which is related to the $\kappa$, the full width half maxima (FWHM) of each nucleation curve, and consistent with Fig. 6d. The parameter, $\chi$, in eqn. 22 defines the nucleation centres, determined by $\theta(z,n)$. The co-efficient $a_1$ and $a_2$ are fitting parameters. This calculation is based on the formulations for the potential profile (Eqns. 13–16) and the corresponding electron transfer rate derived using Marcus theory (Eqns. 17–20).

From the $Nr_{xy}(z,n)$ values, the volumetric fraction of 22, $f_{22}(n)$, is calculated by summing $Nr_{xy}(z,n)$ over z as shown in eqn. 23-24 (see Fig. 8d).

$$f_{22}(n) = \frac{1}{30}\sum_{z=1}^{z=30} Nr_{xy}(z,n) = \frac{1}{30}\sum_{z=1}^{z=30}\left[\frac{1}{\pi} \times \arctan\left(\frac{n-\chi(z)}{\kappa}\right) + \frac{1}{2}\right] \quad (23)$$

$$f_{31}(n) = 1 - f_{22}(n) \quad (24)$$

The optimal slope and spacing between the $Nr_{xy}(z,n)$ curves, governed by $\kappa$ and $\chi(z)$ (eqn. 23), ensure the alignment of their linear regimes. Consequently, we observed linear variation of $f_{22}(n)$ with the conductance level (or pulse count $n$), as demonstrated for the 900 mV, 80 ns pulses. As illustrated in Fig. 8d, the calculated $f_{22}(n)$ is consistent with the experimental data obtained by analysing *in-situ* Raman spectra measured as a function of applied pulses. Fig. 8e-f presents a schematic illustrating the molecular transition within the film from the 31-state to the 22-state, both along the $z$-axis and $x - y$ plane. To experimentally verify the illustration in Fig. 8f, we characterised the local, spatially resolved conductance of the film with a conductive measurement tip at different stages of potentiation[62]. Fig. 9 shows spatial current distribution map that imaged the nucleation process along the $x - y$ plane, supporting the illustration in Fig. 8f.

Thus, to determine the electronic states of a particular molecule within the film, we need to consider both the potential profile along $z$ (Fig. 8a) and the nucleation profiles across $x - y$ (Fig. 8c). The screening of film molecules, influenced by the fraction of constituent electronic states, governs the nucleation centres along the $z$-direction, while nucleation processes at specific $z$-values determine the corresponding areal distribution. From the fraction of different molecular states ($f_{22/31}$) at each conductance state ($n$), obtained using eqn. 23-24 (Fig. 8d), we calculated the fractional change ($df_{22/31}$) and plugged it in the eqn. 10-12, to determine the coupling strengths ($\gamma_m, \Gamma_{L/R}$) and the rate constants ($K_{m,f/b}, K_{L/R,f/b}$) at different conductance levels in the eqn. 1-6.

The coupling constants in quasi-DC measurements can be expressed using equations 25–27, which share the same structure as equations 10–12, used for pulsed measurements. In both scenarios, the effective coupling constants are represented as a weighted sum of coupling strengths across various redox states. These weights correspond to the fractional occupancy of each redox state, determined by the number of pulses in equations 10–12, whereas in equations 25–27 (for steady state), they are governed by the applied voltage value. As we described in Fig. 2, 3, and 4, while the pulsed measurements (eqn. 10-12) involve only two states, *viz.* 31 and 22, the quasistatic measurements (eqn. 25-27) involve three redox states, 31, 11 and 00.

$$\gamma_m(V) = [f_{31}(V) \times \gamma_{m,31}] + [f_{11}(V) \times \gamma_{m,11}] + [f_{00}(V) \times \gamma_{m,00}] \qquad (25)$$

$$\Gamma_T(V) = [f_{31}(V) \times \Gamma_{T,31}] + [f_{11}(V) \times \Gamma_{T,11}] + [f_{00}(V) \times \Gamma_{T,00}] \qquad (26)$$

$$\Gamma_B(V) = [f_{31}(V) \times \Gamma_{B,31}] + [f_{11}(V) \times \Gamma_{B,11}] + [f_{00}(V) \times \Gamma_{B,00}] \qquad (27)$$

Here, $\gamma_{31/11/00}(V)$ is the inter-molecular coupling and $\Gamma_{T/B,31/11/00}(V)$ is the Top/Bottom electrode-molecule coupling in 31, 11 and 00 state respectively in steady-state. $f_{31/11/00}(V)$ is the fraction of the molecules in 31, 11, and 00 state respectively at different measured voltage.

Using these rate constants for pulsed as well as steady state measurements, we can solve for the occupational probabilities ($P_i$) from eqn. 7-8 and calculate the current ($I(n)$) from eqn. 9, which can be expressed in a closed form as presented in the eqn. 28

$$I(n, V_r) = e \times \gamma_m(n) \left( \exp\left[\frac{(V_r')}{\eta k_B T}\right] \times P_{N-1}(n, K_{m,f/b}, K_{T,f/b}, K_{B,f/b}) - \exp\left[\frac{-V_r'}{\eta k_B T}\right] \times P_N(n, K_{m,f/b}, K_{T,f/b}, K_{B,f/b}) \right) \qquad (28)$$

Here, $V_r' = a_m \times V_r$, $P_{N-1}$ and $P_N$ is obtained by solving eqn. 7-8.

As illustrated in Fig. 10, the computed current responses could reproduce experimental potentiation-depression characteristics as well as the steady-state $I(V)$ behaviour. In Fig. 10a, the calculated linear and symmetric weight update curve, derived from equations (1–24), closely matches the experimental data. The inset presents semi-log plots to highlight the model's ability to capture the accuracy across full dynamic range of conductance. Fig. 10b shows that as the number of pulses exceeds 16,500, the potentiation curve becomes non-

linear—a phenomenon accurately reproduced by our model. This deviation arises from the dominance of the non-linear regimes of the nucleation curves. Similarly, in Fig. 10c, for higher pulse amplitudes (1.22 V), the overlapping non-linear nucleation regimes are well-captured by our formulation, yielding a strong match with experimental traces. In Fig. 10d, we show the correspondence between experimental and calculated quasi-DC $I(V)$ traces. Fig. 11a summarizes how our formulation not only reproduces linear and symmetric characteristics but also successfully models deviations from linearity. Fig. 11b demonstrates that even when pulsing is halted before 16,500 cycles and depression is initiated, the system still exhibits linear and symmetric responses, consistent with experimental observations.

Notably, although we did not provide here a detailed description of the fabrication process[63-66], we emphasize that the quality of the interface plays a pivotal role in preserving linearity. To ensure linearity, the applied voltage must predominantly drop across the molecular film. Any significant voltage drop across the electrode-film interfaces can compromise this linearity. In Figure 12, we illustrate that poor interface quality, resulting in 60% of the applied voltage dropping across the interfaces, disrupts the zeroth-order nature of the molecular transition (Figure 12a). This, in turn, causes the potentiation curve to deviate substantially from linearity (Figure 12b).

The objective of our mathematical framework was to develop and refine a parameter space to achieve linear and symmetric weight update behaviour. Figure 13 illustrates the impact of key parameters on linearity and symmetry, presenting a flowchart contrasting two scenarios: one resulting in linear weight update characteristics (right panel, Fig. 13f–i) and the other exhibiting non-linearity (left panel, Fig. 13a–e). The three critical parameters influencing linearity are:

A. **Nucleation rate which is inversely proportional to the FWHM of each nucleation curve ($\kappa^{-1}$).** A lower kinetic barrier or higher attempt frequency of molecular transition would result in a higher nucleation rate, and thus, a lower FWHM of the nucleation curve (steeper) introducing saturation and non-linearity. This explains why transitions driven by ionic movements are better suited for achieving linear responses. Ionic movements are associated with a larger reduced mass, lower attempt frequencies, and consequently lower nucleation rate. These characteristics allow for operation within closely packed energy pockets separated by smaller barriers.
B. **$z$-gradient of nucleation ($a_1 = d\chi(z)/dz$) which is inversely proportional to the applied pulse amplitude**. Its value is determined by the device's screening ability at various conductance levels. Higher pulse amplitudes cause more molecules to transition into the 22-state with each pulse, which reduces the film's screening capacity between successive conductance states. Consequently, nucleation curves converge more closely at higher pulse amplitudes, leading to a lower $d\chi(z)/dz$ ($a_1$ in equation 22), as shown in Fig. 13b. Hence, for linearity lower amplitude pulses that can still drive the thermodynamic transition is preferred.

Linear behaviour is maintained only within a small perturbation regime with an optimal value of $\kappa$ and the pulse height (see Fig. 13f,g).

C. **The ratio of the coupling constants ($\gamma_{22}/\gamma_{31}$)** which describes the change in molecular coupling strength as the device transitions from the 31-state to the 22-state. A higher value of $\gamma_{22}/\gamma_{31}$ facilitates linearity across a broader dynamic range of conductance, allowing for a greater number of conductance levels and improved analog resolution (Fig. 13h-i).

Based on this parametric analysis, we constructed a phase diagram in Fig. 14 to visualize the effect of these parameters on linearity. We introduced a linearity factor for potentiation ($v_P$) and depression ($v_D$) as given in Equation 29-32. The values close to 0 signifies perfect linearity, while larger values represent maximum deviation from linearity.

$$\nu_P = \sum_{n=1}^{n_{max,P}}[I(n) - I(n-1) - \mu_P]^2 / \mu_P \qquad (29)$$

$$\mu_P = [I(n_{max,P}) - I(1)]/n_{max,P} \qquad (30)$$

$$\nu_D = \sum_{n=1}^{n_{max,D}}[I(n-1) - I(n) - \mu_D]^2 / \mu_D \qquad (31)$$

$$\mu_D = [I(1) - I(n_{max,D})]/n_{max,D} \qquad (32)$$

$n_{max,P}$ and $n_{max,D}$ are the total number of pulses applied during potentiation and depression respectively. $I(n)$ is calculated using eqn. 9.

The phase diagram (Fig. 14) reveals optimal conditions for linearity, including a lower nucleation rate ($\kappa^{-1}$) in the range of (0.016 – 0.05) µs$^{-1}$ for potentiation and (0.031 – 0.055) µs$^{-1}$ for depression. Additionally, a lower rate of change in screening capacity, driven by smaller pulse amplitudes (900–1000 mV for potentiation and 750–892 mV for depression, indicated by $(\frac{d\chi(z)}{dz})^{-1}$, confines the system to a low-perturbation regime ensuring linearity in weight update responses. Furthermore, a large $\gamma_{22}/\gamma_{31}$ value ensures linearity over a wide dynamic conductance range (see eqn. 10), in our case, extending up to four orders of magnitude, allowing for tens of thousands of conductance levels.

In closing, we highlight the five key takeaways from this work:

First, the realization of nearly all essential neuromorphic attributes in our molecular circuit elements is enabled by the culmination of molecular design and precise fabrication, honed over nearly a decade of experimentation that could provide transformative insights into designing next-generation neuromorphic materials. Historically, the field has been dominated by filamentary mechanisms, which, being inherently stochastic, have lacked consistency. While individual devices might achieve high accuracy, scaling this precision to crossbars or circuit-level analog systems remains exceptionally challenging. A more promising alternative lies in uniform thermodynamic transitions governed by kinetics—mechanisms that are not only more controllable but also volumetrically uniform, enabling significantly higher densities of switchable units and thus, inherently larger number of accessible states.

Second, electron transport in molecular solids is governed by many-body physics, offering unique design opportunities to tailor functionalities. For example, in our molecular films, ions can occupy energetically degenerate pockets around molecules. These varying configurations produce distinct transfer functions for electron transport, potentially unlocking a vast range of conductance states. While this dimension was only partially explored in this study, it shows that there is a plenty of room for future research that could further finetune and enhance the performance.

Third, achieving the desired electronic responses from new materials often requires drawing upon concepts from diverse scientific disciplines. In this work, we utilized zeroth-order kinetics — a principle well-established in molecular materials for over a century — which proved ideal for achieving the linearity and symmetry critical for neuromorphic applications.

Fourth, an optimal fabrication process is crucial. Significant potential drops across interfacial barriers can disrupt molecular transitions and introduce non-linearity. Therefore, achieving high-quality molecule-film contacts is essential to obtain the desired kinetic control.

Finally, and perhaps most significantly, we demonstrated that even inherently nonlinear mechanisms, such as electrostatic screening and nucleation, can deliver near-ideal linear and symmetric responses when carefully constrained within small perturbation regimes. This approach offers broad applicability across molecular systems, oxides, and ferroelectrics. The ultimate challenge for materials scientists will be mastering the precise orchestration of local nonlinearities, a task that demands innovation tailored to each specific material system.

**Table 2 –** List of modelling parameters used in eqn. 1-28

| Symbol | Description |
|---|---|
| $K_{m,f/b}$ | Forward/backward intermolecular electron transfer rates |
| $K_{T,f/b}$ | Forward/backward electron transfer rates between the top electrodes and the molecular layer |
| $K_{B,f/b}$ | Forward/backward electron transfer rates between the bottom electrodes and the molecular layer |
| $\gamma_m$ | Molecule-molecule coupling constant |
| $\Gamma_{T/B}$ | Top/bottom electrode-molecule coupling constant |
| $\gamma_{m,22/31}$ | Molecule-molecule coupling constant in 22/31-state |
| $\Gamma_{T/B,22/31}$ | Top/bottom electrode-molecule coupling constant in 22/31-state |
| $a_m$ | Fraction of applied voltage drops across molecule |
| $a_{T/B}$ | Fraction of applied voltage drops across top/bottom electrode |
| $\Delta E$ | Hopping energy gain |
| $u_\lambda$ | Interfacial re-organization energy |
| $V_r$ | Reading pulse amplitude |
| $\eta$ | Fitting parameter in eqn. 1-2 |
| $P_i$ | Occupational probabilities at the different hopping sites |
| $N_z$ | Total number of hopping sites along z |
| $n$ | Different analog conductance states |
| $df_{22/31}$ | Fractional change in the 22/31-state between two consecutive conductance states |
| $I$ | Current in different conductance states |
| $\Phi_{mol}$ | Potential profiles along z |
| $\zeta$ | Screening parameter |
| $V_b$ | Writing pulse amplitude |
| $k_{ET}$ | Electron transfer rate between 3 and 1 |
| $\gamma_{DA}$ | Coupling between the 3 and 1 |
| $P$ | Probability of electron transfer between 3 and 1 |
| $\Delta G^0$ | Gibbs free energy for the 31→22 transition |
| $\varepsilon$ | Stochastic fitting parameter in eqn. 20 |
| $Nr_{xy}$ | Population of 22 along $x$-$y$ planes at different $z$-values |
| $\chi$ | Nucleation centres for layers at different $z$-values |
| $\kappa$ | Rate of the nucleation processes |
| $a_1$ and $a_2$ | Parameters derived from eqn. 13-20, used for fitting in eqn. 22 |
| $f_{22/31}$ | Volumetric fraction of 22/31-state |
| $\nu_{P/D}$ | Ideality factor for potentiation/depression |
| $\mu_{P/D}$ | Conductance difference between any two adjacent levels for potentiation/depression in case of linear response |

**Acknowledgements:** This research was supported by the IISc Startup Grant, Pratiksha Trust Grant, SERB Core Research Grant (CRG/2022/001998), and the MeitY Project, *"Neuromorphic Molecular Nanotechnology Enabled CMOS Chip for AI on the Edge"* (Award No. Y-21(1)/2024-R&D-E). SG expresses gratitude to Prof. Sreebrata Goswami and Damien Thompson for their valuable insights. BK acknowledges the support of the Prime Minister's Research Fellowship (PMRF). We thank Agnès Tempez for the c-AFM and TERS measurements and related discussions.

**Data availability:** The data that support the findings of this study are available from the corresponding authors on reasonable request.

**Author Contributions:**

SG conceived and designed the project. BK collected data. SG, BK analysed the data. SG and BK performed the calculations. SG and BK wrote the paper.

**Figures:**

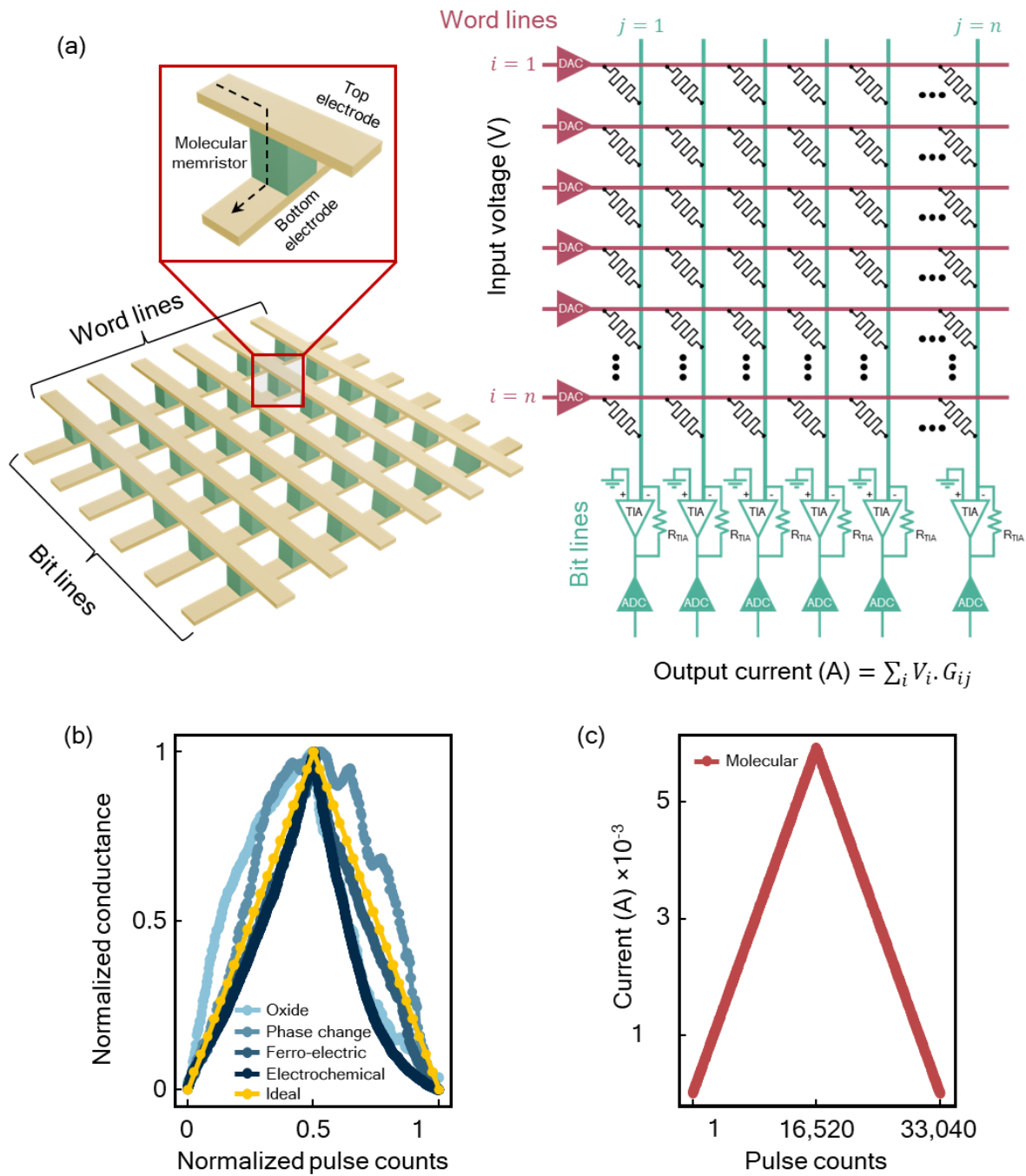

**Figure 1: Crossbar Structure and Desired Electrical Characteristics** – (a) A schematic of a molecular crossbar with a magnified view of a single crosspoint, illustrating the direction of current flow. In this architecture, the conductance values at the crosspoints represent the encoded matrix, while the applied voltages correspond to the input vector, enabling vector-matrix multiplication in a single computational step. (b) A comparative analysis of potentiation and depression characteristics across memristors based on various material systems, including oxides, phase-change materials, ferroelectrics, and electrochemical systems. The x-axis is normalized by the total number of pulses used in the measurement, and the y-axis is scaled between the minimum and maximum conductance values achieved. Data are sourced from references[10,18,21,27]. (c) Potentiation and depression characteristics recorded in the molecular memristor developed in this work.

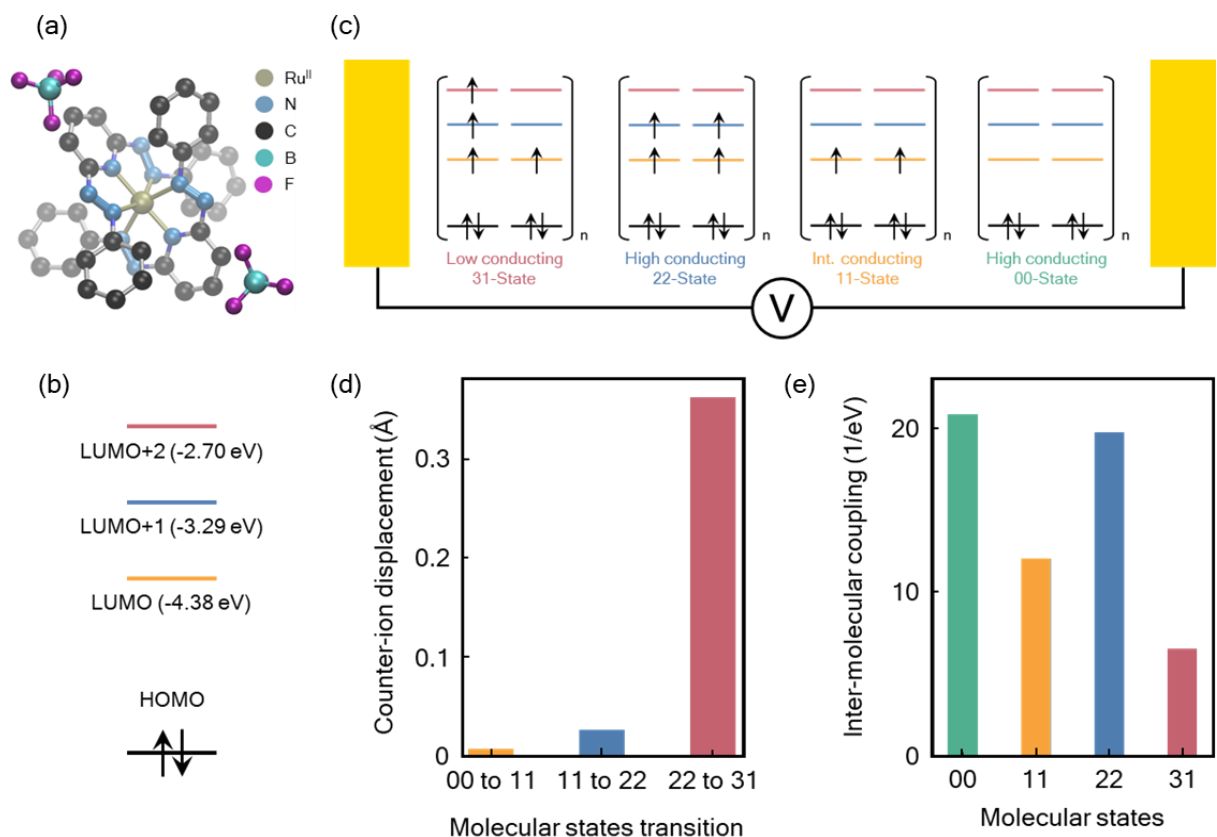

**Figure 2: Molecular States in the Device** – (a) Molecular structure of $[Ru^{II}L_2](BF_4)_2$, where L represents 2,6-bis(phenylazo)pyridine. (b) Molecular orbital (MO) energy level diagram of $[Ru^{II}L_2]^{2+}$, highlighting the HOMO (highest-occupied molecular orbital) and LUMO + n (lowest-unoccupied molecular orbital) energy levels (n is an integer ranging from 0 to 2). (c) Schematic representation of experimentally observed electronic states — 31, 22, 11, 00 — at different applied voltages. Here, "0" represents the unreduced molecular state, while "1," "2," and "3" correspond to the one-, two-, and three-electron reduced states, respectively. The occupancy of anti-bonding orbitals in a pair of molecules is shown to illustrate symmetry breaking in the 31 state. (d) Calculated displacement of counter-ions during transitions between molecular states. (e) Calculated inter-molecular coupling for each molecular state using methods illustrated in reference[64].

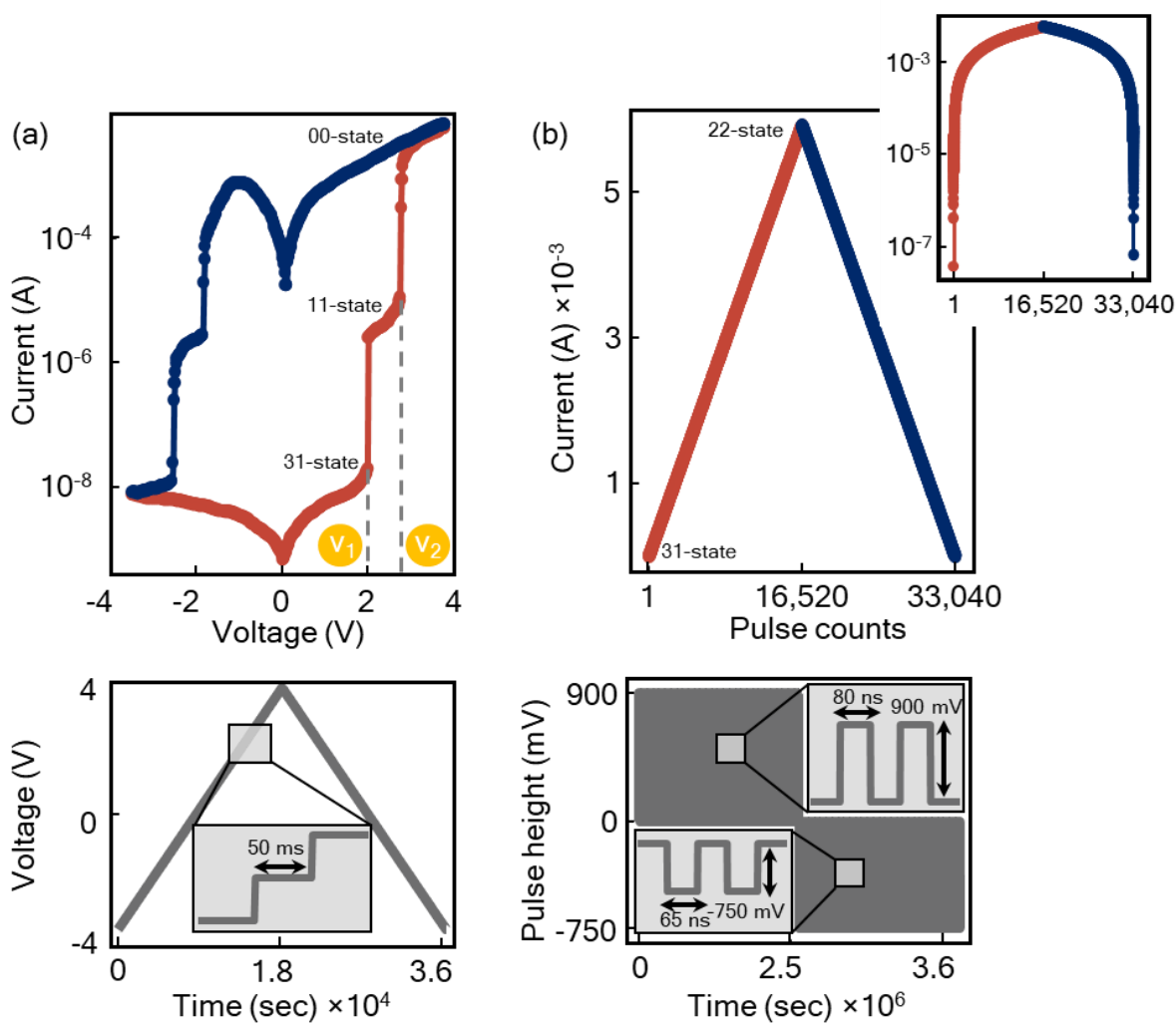

**Figure 3: Digital vs. Analog Response in the Molecular Memristor** – (a) Current-voltage (I-V) characteristics from memristors made with Ru$^{II}$L$_2$](BF$_4$)$_2$ showcasing digital switching with sharp, discrete transitions. These correspond to the transitions from the 31 to 11 state and the 11 to 00 state (see Fig. 2c) at voltage thresholds V$_1$ and V$_2$, respectively. The lower panel illustrates the time-dependent voltage input used during the I-V measurements. The time duration of each current measurement is about 50 ms for individual voltage points, as detailed in the inset. (b) Current response to sub 100ns voltage pulses demonstrating analog switching, characterized by linear and symmetric potentiation and depression. In this mode, molecular transitions occur between the low-conducting 31 state and the high-conducting 22 state via inter-molecular electron transfer (as substantiated later in Fig. 4). The inset shows a semi-log plot highlighting the preserved linearity and symmetry across a conductance range spanning more than four orders of magnitude. The bottom panel depicts the time-dependent pulse input used during these measurements. Potentiation was induced using +900 mV pulses of 80 ns width, while symmetric depression was achieved with −750 mV pulses of 65 ns width, as detailed in the inset.

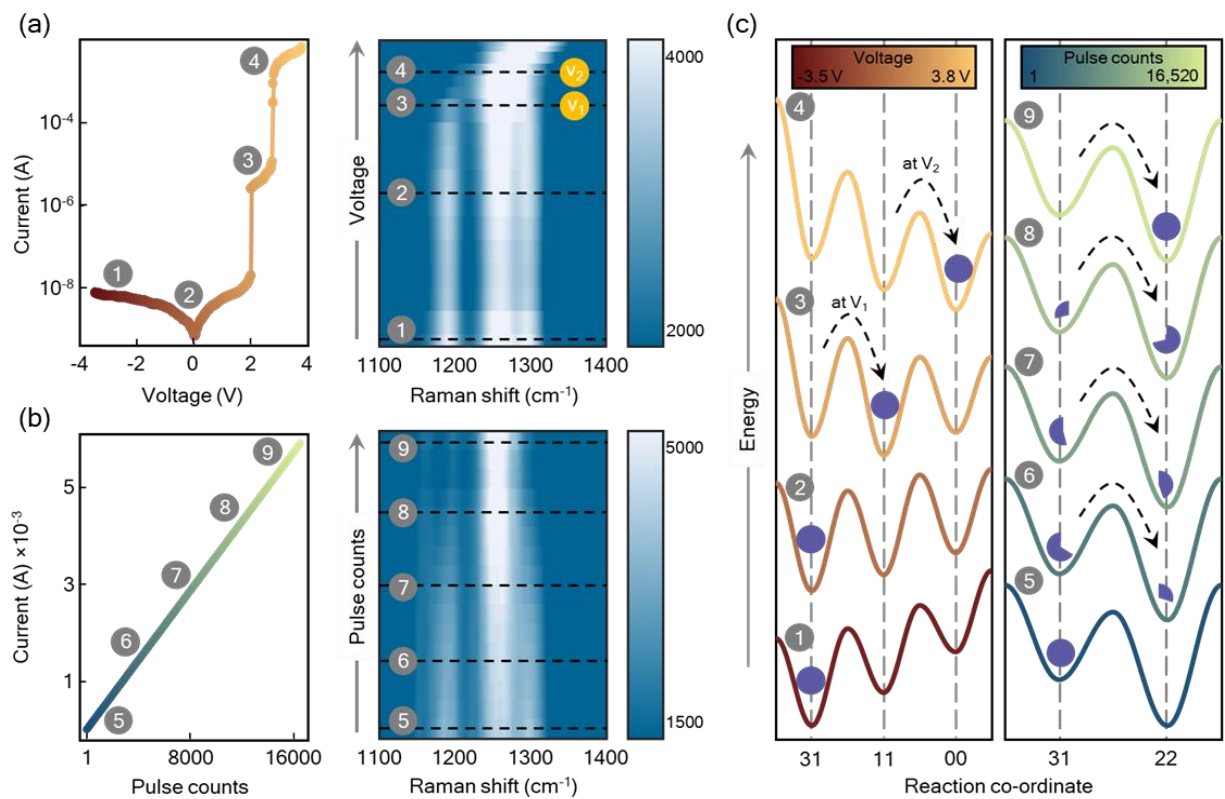

**Figure 4: In-Situ Raman Spectroscopy Analysis** – (a) Current-voltage (I-V) characteristics during a forward voltage sweep in digital switching along with a pseudo-color map of voltage-resolved Raman spectra collected throughout the sweep. (b) Current response to pulsed input during potentiation in analog switching, alongside a pseudo-color map of pulse-resolved Raman spectra recorded during the process. (c) Reaction coordinate diagrams illustrating molecular state transitions. The left panel represents the discrete transitions between the 31, 11, and 00 states during the steady state voltage sweep (Fig. 3a), with points 1–4 corresponding to specific stages indicated in Fig. 4a. The right panel shows the fraction of molecules transitioning from the 31 to 22 states under pulsed input, with points 5–9 correlating to stages identified in Fig. 4b.

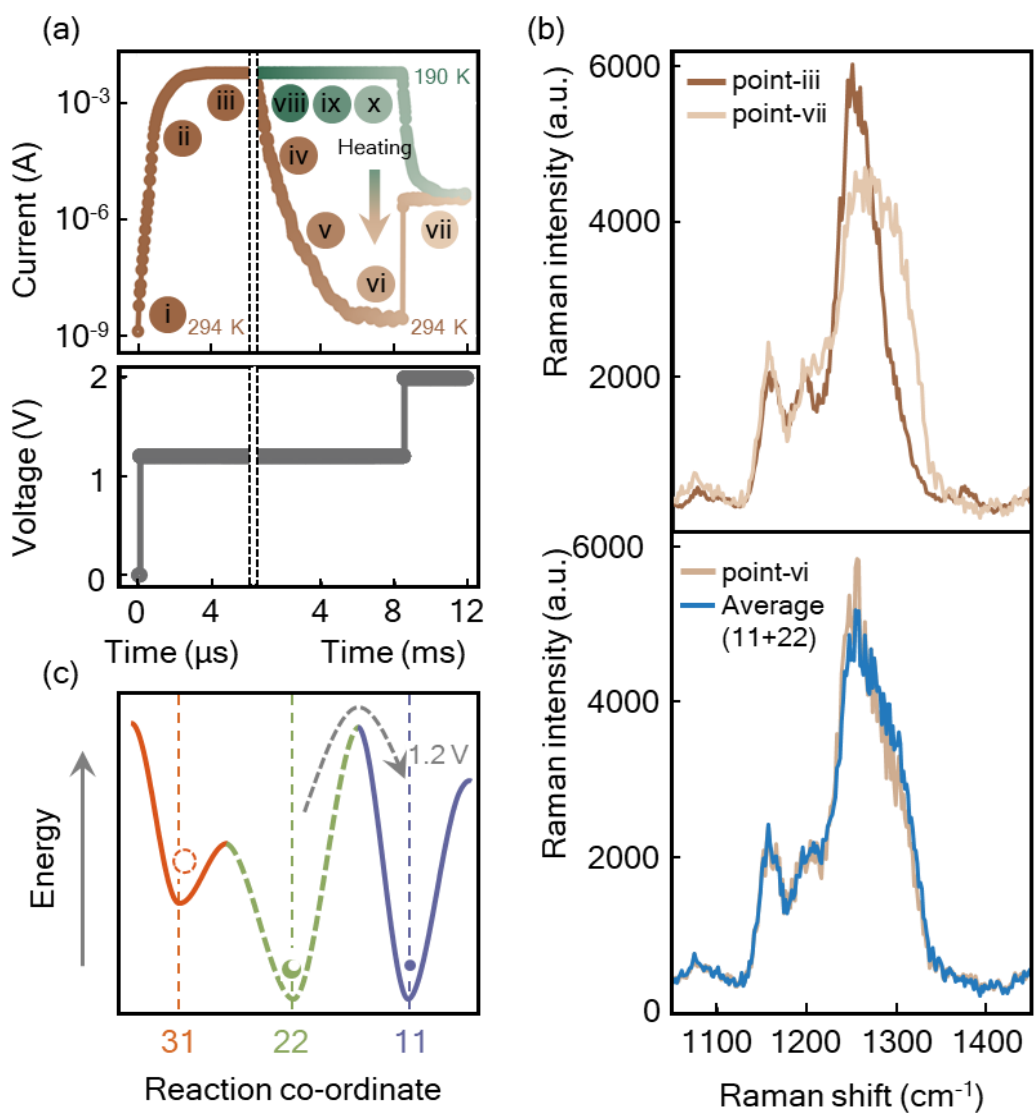

**Figure 5: Temporal Dynamics of the 22-State** – (a) Time-resolved pulsed electrical measurements illustrating transitions between the 31, 22, mixed (11+22), and 11 states at temperatures of 294 K and 190 K. (b) Raman spectra corresponding to the pure 11 and 22 states, as well as the mixed state, recorded at points iii, vii, and vi, respectively. The Raman spectrum recorded at point vi closely aligns with a linear combination of the pure 11 and 22 state spectra, confirming an equal mixture of the two states. This observation suggests that the 11 and 22 states are nearly degenerate, causing the 22 state formed at point iii to evolve into a (11+22) mixture over time. (c) Reaction coordinate diagram depicting the transitions between various electronic states of the molecular system.

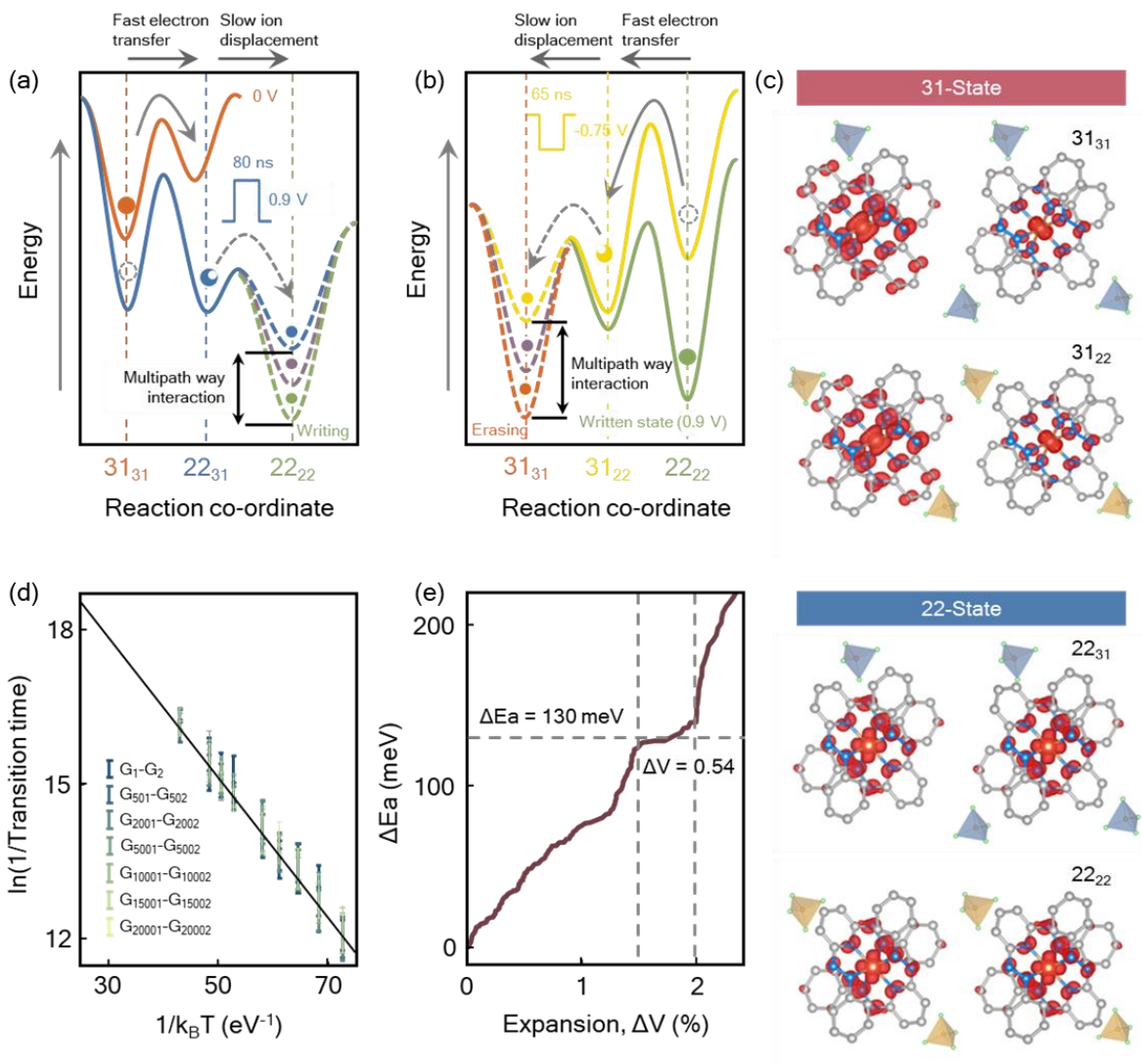

**Figure 6: Electron Transfer vs. Counter-Ion Displacement** – (a, b) Reaction coordinate diagrams depicting the dynamics of fast electron transfer followed by slow counter-ion displacement during (a) potentiation and (b) depression processes. (c) The four electronic and ionic configurations derived from the reaction coordinate diagrams in (a) and (b). The electron densities in the antibonding orbitals for the 31 and 22 states are calculated from quantum chemical calculations[3]. (d) Transition barriers between successive states (see legend), determined from temperature-dependent transition rate measurements conducted between 290 K and 160 K. Error bars represent ±3σ (σ = standard deviation), based on seven independent measurements. The calculated activation barrier for transitions between different levels is consistent (~134 meV). (e) Volumetric expansion resulting from counter-ion relaxation at approximately 130 mV, corroborating the activation barrier deduced in (d).

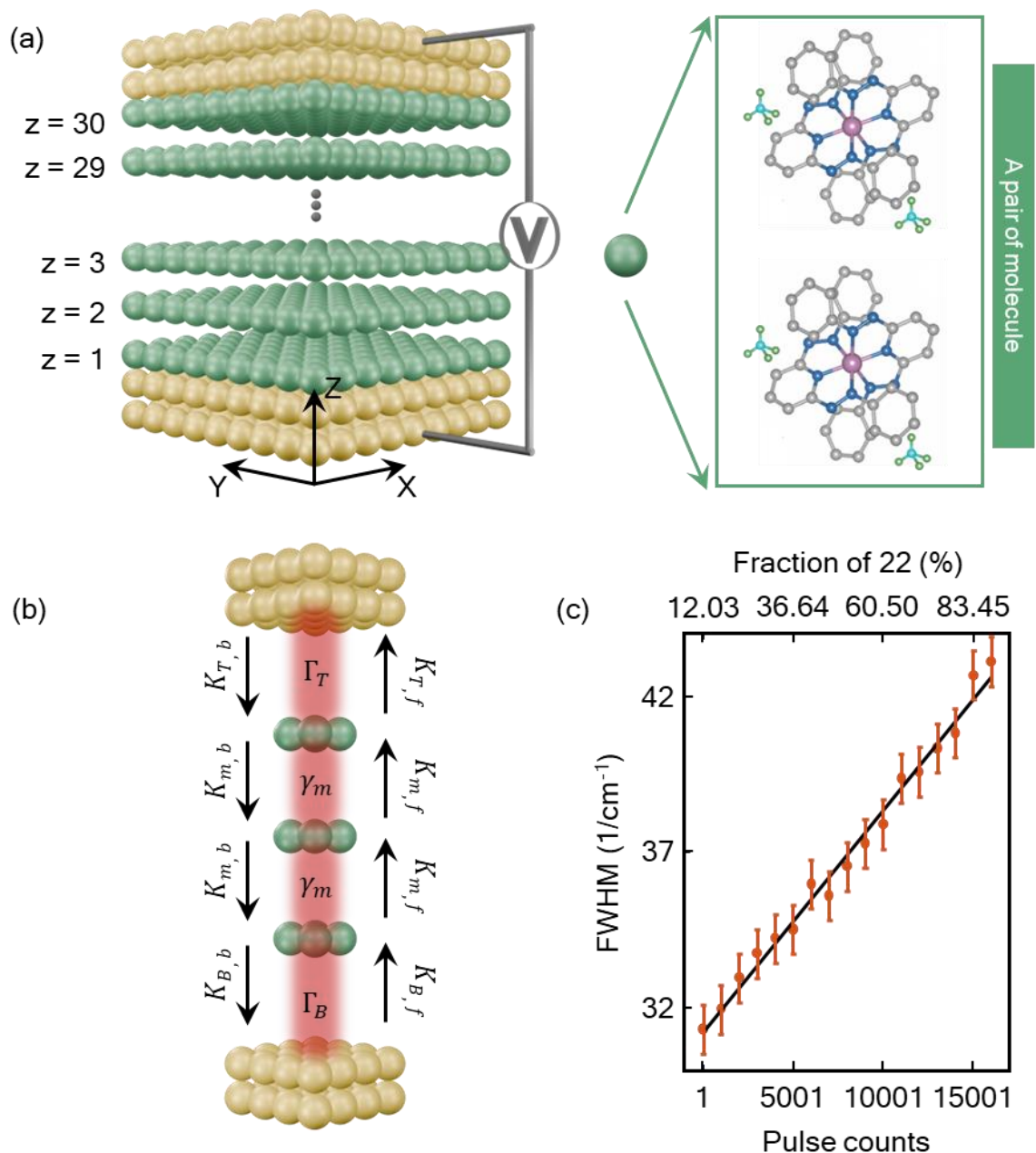

**Figure 7: Modeling Charge Transport in the Molecular Device** – (a) Schematic representation of the molecular device, illustrating a molecular film sandwiched between the top and bottom electrodes. The film is abstracted as molecular layers layers along the z-axis, where each 2 nm-thick layer comprises a pair of molecules. The complete molecular film spans approximately 60 nm, encompassing 30 such layers. (b) The coupling parameters and rate constants incorporated into our current transport formulations, see equations 1–6. (c) A linear variation in the full-width half-maximum (FWHM) of azo Raman modes (hopping orbitals), reflecting a proportional increase in molecular coupling strength with the number of applied pulses, evidenced by their progressive broadening.

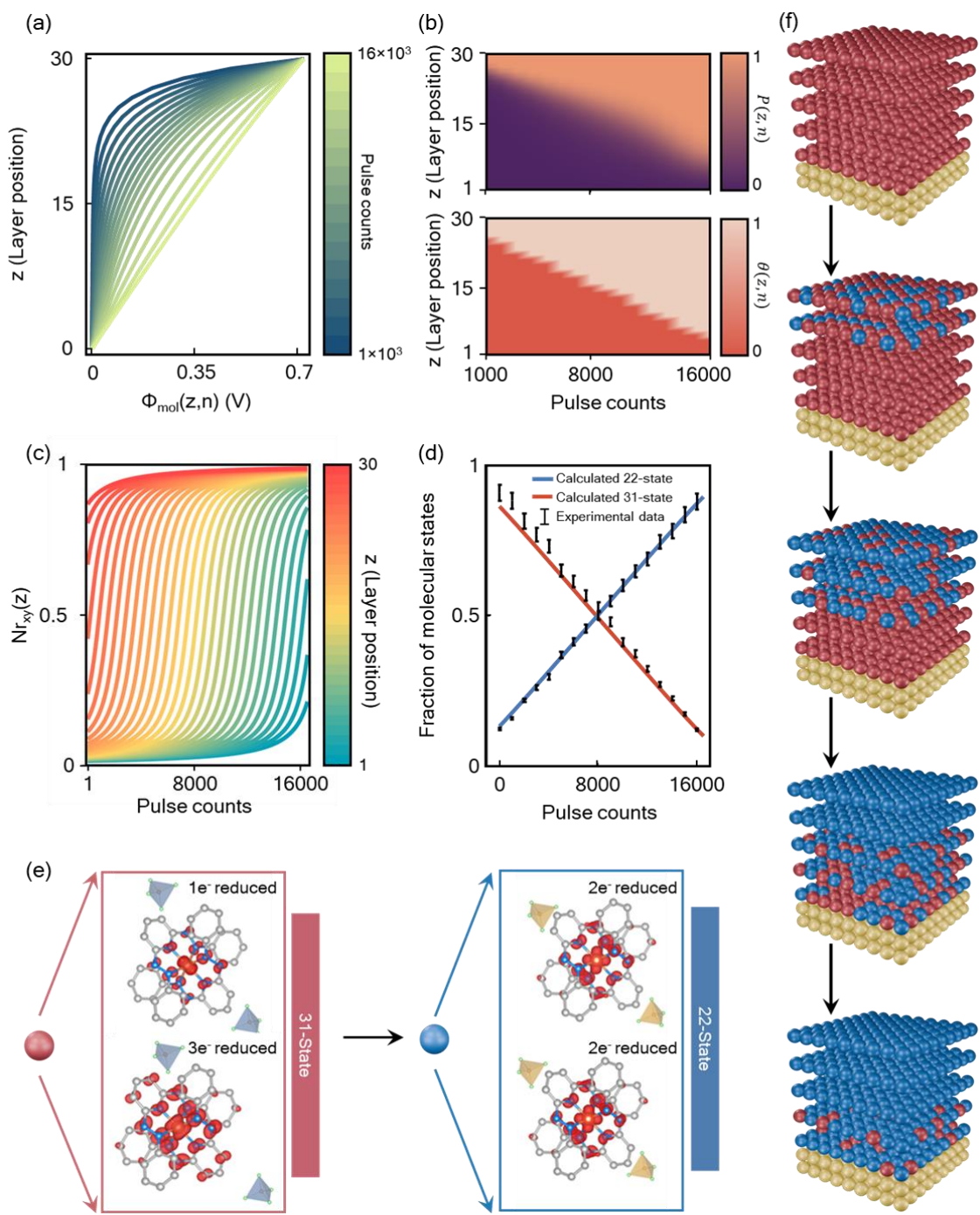

**Figure 8: Modeling the Evolution of the 22-State** – (a) Calculated potential profiles at different pulse intervals, showing how the potential landscape evolves with pulse count. (b) Top panel: Electron transfer probability between the 3 and 1 states, $P(z,n)$, computed using equation 19. Bottom panel: $\theta(z,n)$, calculated using equation 20 (c) Nucleation curves $Nr_{xy}(z)$ calculated from equations 21–22. (d) Comparison of calculated fractions of 22 and 31 states across 16,520 pulses, using equations 23–24, with experimental data overlaid. (e) Representation of the 31 and 22 states. (f) Schematic depicting the gradual volumetric evolution of the 22-state within the device.

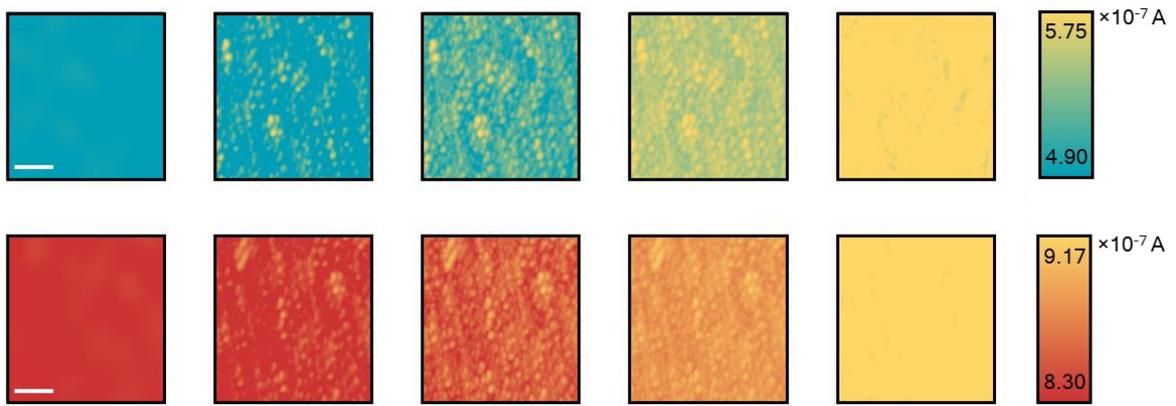

**Figure 9: Evolution of the 22-State in c-AFM Measurements** – c-AFM scans showing the progression of the 22-state with incremental pulse applications. Two distinct current switching regimes are presented to highlight the consistency of the switching mechanism. The scale bar represents 500 nm.

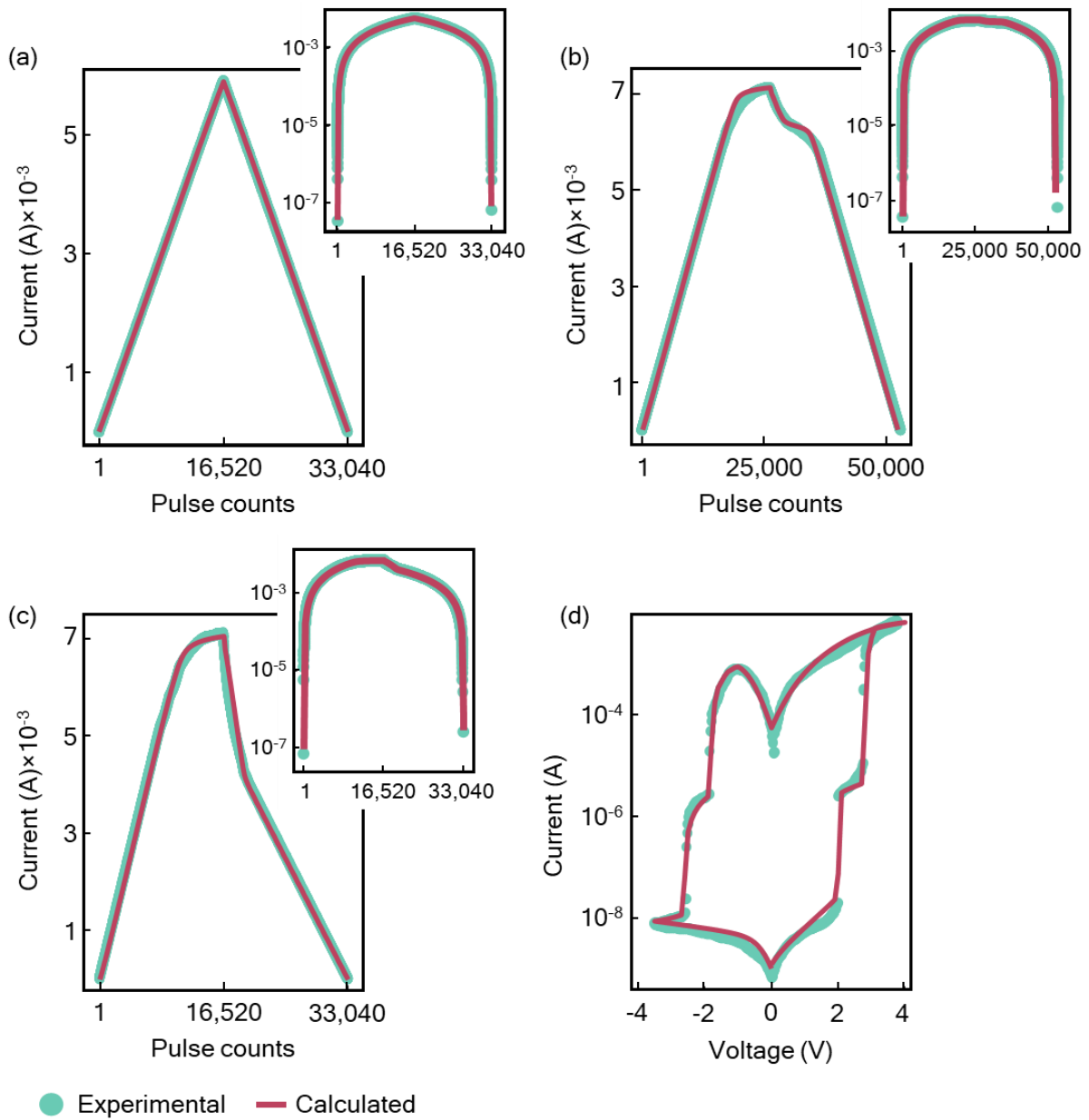

**Figure 10: Comparison of Modeling Results with Experimental Data** – (a-c) Fitted current versus pulse data for analog switching: (a) in the linear regime with 900 mV writing pulses, (b) in the non-linear regime beyond 16,520 number of 900 mV pulses, and (c) at a higher pulse amplitude of 1.22 V. Logarithmic scale plots are provided in the insets. (d) Fitted current versus voltage data for digital switching.

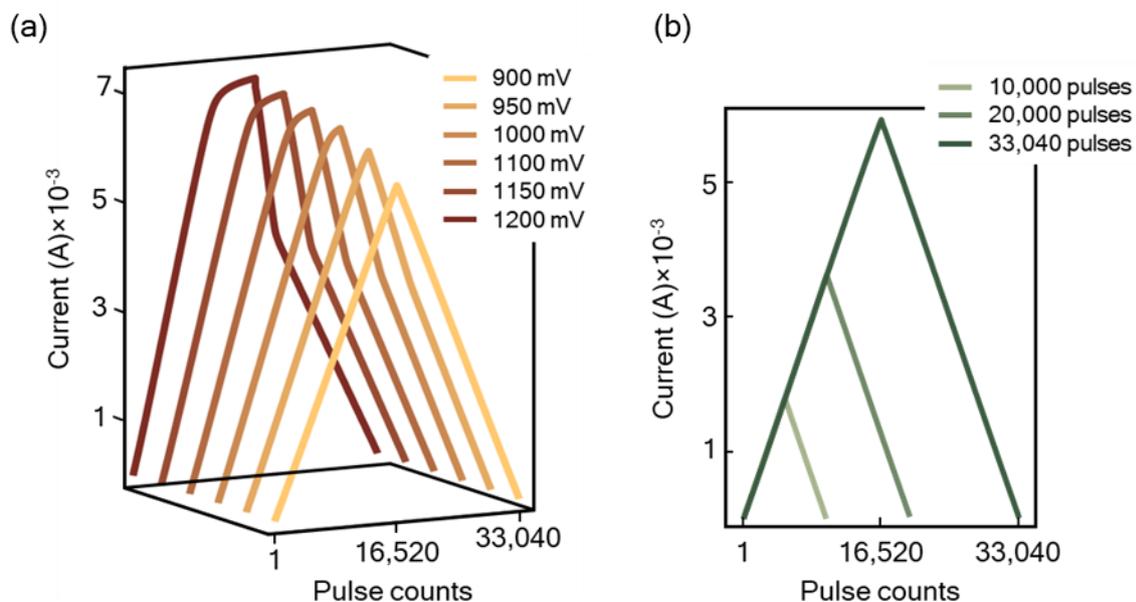

**Figure 11: Further Modeling Outcomes** – Predicted potentiation and depression characteristics using the same model framework: (a) for various pulse amplitudes ranging from 900 mV to 1220 mV, highlighting their effect on linearity and symmetry, and (b) for different numbers of pulses within the linear regime.

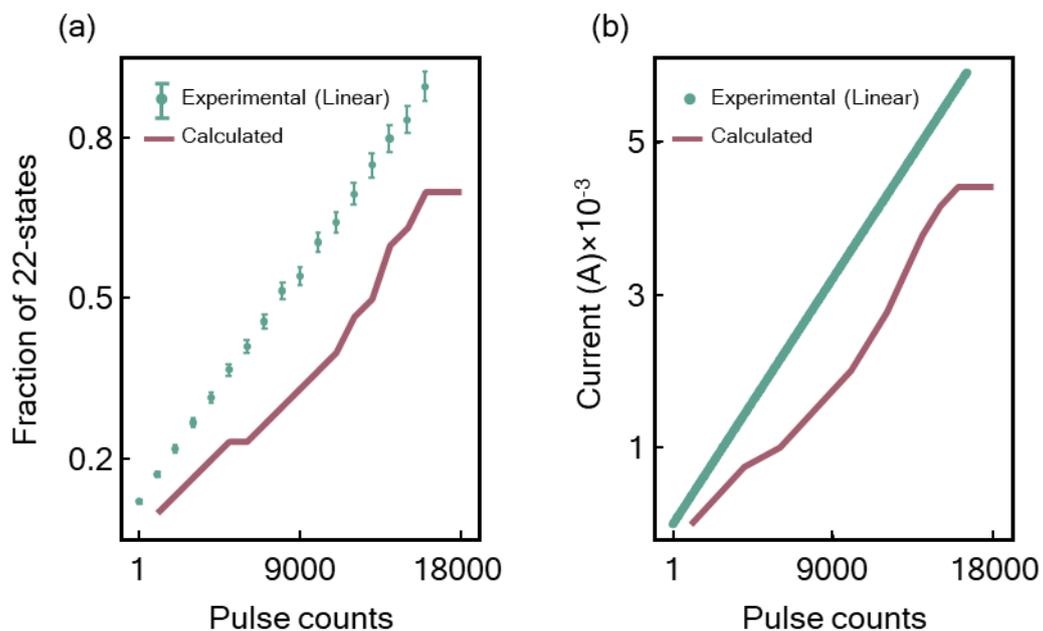

**Figure 12: Effect of interface** – The effect of the interface on (a) molecular transition to 22-state and (b) current, with $\alpha_T + \alpha_B \sim 0.6$. Calculations were performed at intervals of 1000 pulses using equations 1-20 and compared with the ideal linear response.

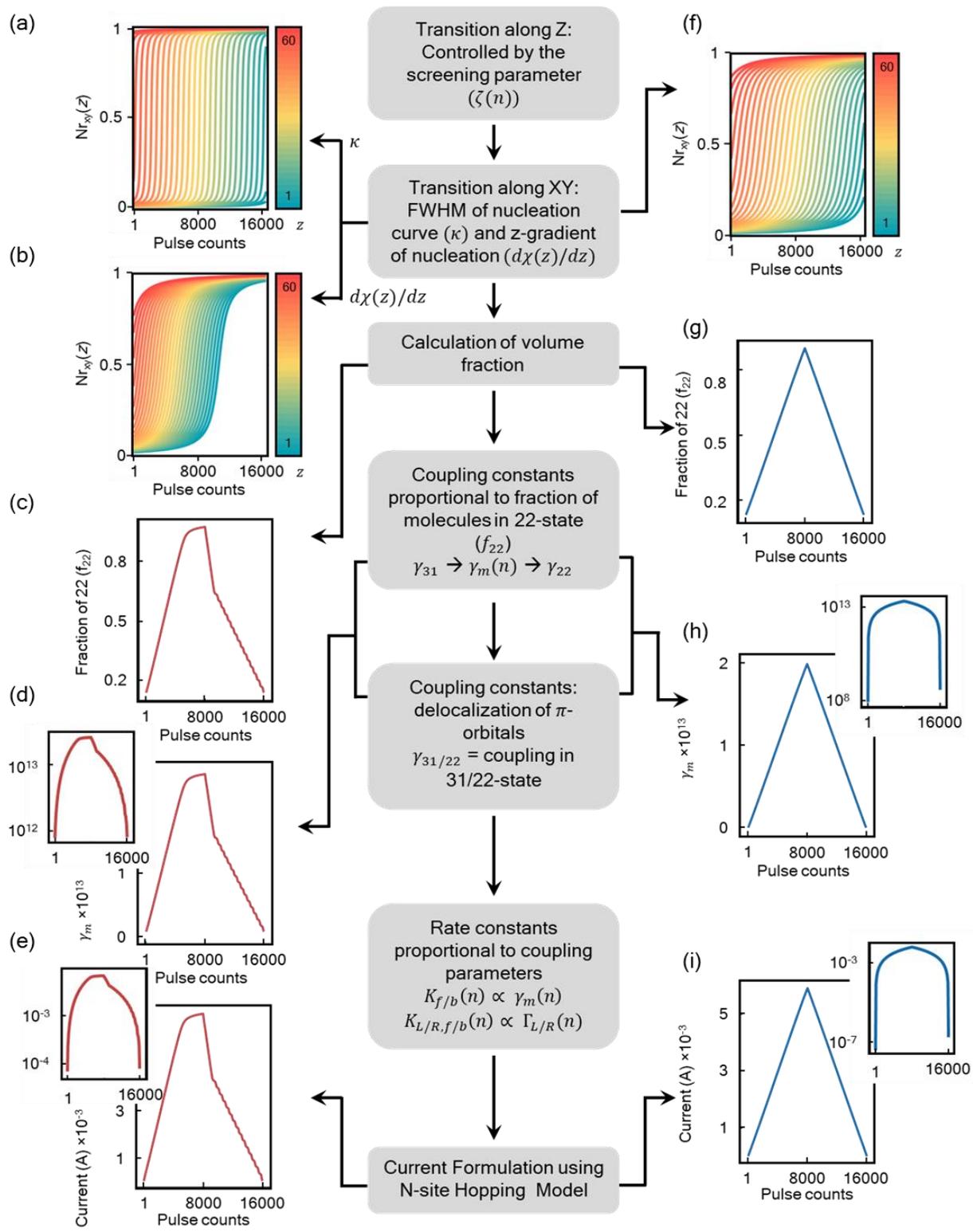

**Figure 13: Explaining the Parameters for the Linear Regime –** Schematic illustrating the steps in current calculation and the impact of key parameters controlling both: (a-e) the non-linear regimes and (f-j) the linear regime.

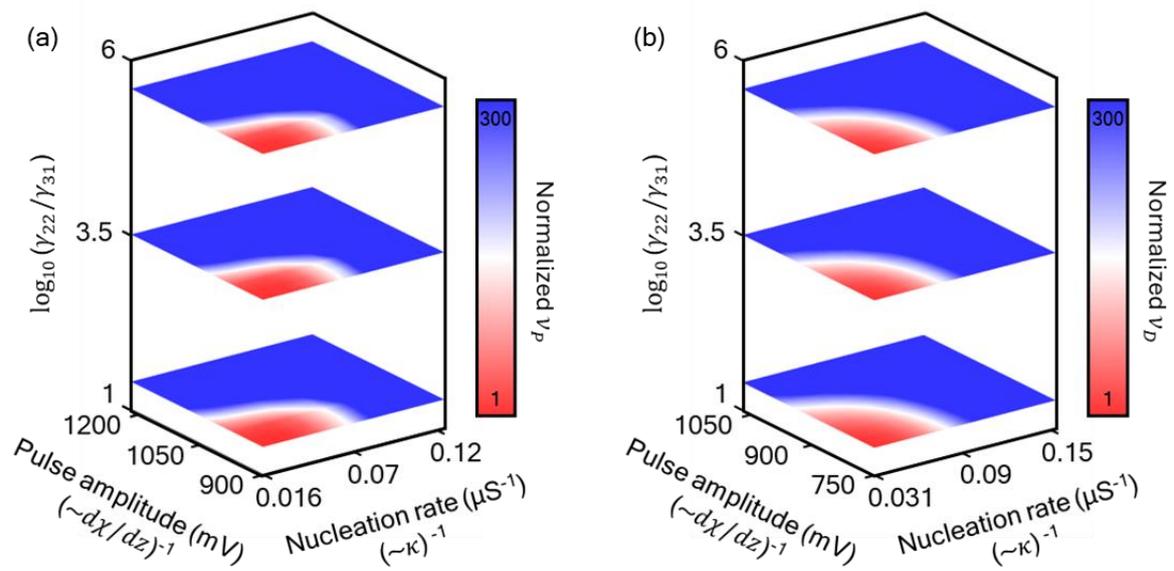

**Figure 14: Parameter Space Controlling Linearity and Non-linearity** – Visualization of the impact of various parameters on the linearity of (a) potentiation and (b) depression. The linearity factors $\nu_P$ (for potentiation) and $\nu_D$ (for depression) are defined in the equations 29-32 and is normalized relative to their minimum values. The nucleation rate is calculated using a pulse width of 80 ns for potentiation and 65 ns for depression. The key takeaway is that careful coordination of multiple parameters, which limit molecular transitions within a small perturbation regime, facilitates linearity.